\documentclass[aps,prl,showpacs,twocolumn,groupedaddress,amssymb]{revtex4}

\usepackage{graphicx}
\usepackage{dcolumn}
\usepackage{bm}
\usepackage{subfigure}
\usepackage{amssymb}
\usepackage{dcolumn}
                                                                                
\begin{document}
\draft
                                                                                
\title{Coarse-graining the dynamics of coupled oscillators}

\author{Sung Joon Moon$^1$}
\author{R. Ghanem$^2$}
\author{I. G. Kevrekidis$^1$}
\email[]{yannis@arnold.princeton.edu}
                                                                                
\affiliation{$^1$Department of Chemical Engineering \& 
Program in Applied and Computational Mathematics (PACM),\\
Princeton University, Princeton, NJ 08544, USA\\
$^2$Department of Civil Engineering, 
University of Southern California, Los Angeles, CA 90089, USA}

\date{\today}

\begin{abstract}
We present an equation-free computational approach to the study
of the coarse-grained dynamics of {\it finite} assemblies of 
{\it non-identical} coupled oscillators at and near full
synchronization.
We use coarse-grained observables which account for the 
(rapidly developing) correlations between phase angles and oscillator natural
frequencies.  
Exploiting short bursts of appropriately initialized detailed
simulations, we circumvent the derivation of closures for the
long-term dynamics of the assembly statistics.
\end{abstract}


\maketitle
\nobreak

Since Winfree's pioneering work in 1960's~\cite{winfree67}, coupled
oscillator models have been investigated extensively.
Some exact results on the collective dynamics for
an infinite number of coupled oscillators (the so-called
continuum-limit)~\cite{kuramoto75,kuramoto84,ariaratnam01}
have shed light on synchronization phenomena in
biological~\cite{winfree67,kuramoto84,gray89,buck88,walker69,neda00},
chemical~\cite{ertl91,kiss02}, and physical
systems~\cite{wiesenfeld96,oliva01}.
%
However, even in this ideal limit, some basic questions
including global, quantitative stability of asymptotic states,
still remain open~\cite{strogatz00,pikovsky01,mikhailov04,acebron05}.
Many real-world systems consist of a large, {\em finite}
number of {\em non-identical} entities, where statistical techniques
for the continuum-limit are not directly applicable.
Exploring and understanding the dynamics of such {\em finite} 
oscillator assemblies is an important topic
(e.g., see Ref.~\cite{balmforth00}).

We present a computer-assisted approach to modeling the
{\it coarse-grained} dynamics of such large, finite
oscillator assemblies at and near full synchronization.
The premise is that there exist a small number of coarse-grained
variables (observables) adequately describing the long-term
dynamics,
and that a closed evolution equation for these observables
exists, but is not explicitly available.
To account for oscillator variability within the assembly,
we treat both the variable oscillator properties (here, natural
frequencies $\omega$) and the oscillator states
(here, phase angles $\theta$) as random variables.
Recognizing a quick development of correlations between
$\omega$ and $\theta$,
we express the latter as a polynomial expansion of the former
(borrowing Wiener polynomial chaos (PC) tools~\cite{ghanem91});
the PC expansion coefficients are our coarse observables.

Availability of the governing equations for the
variables of interest is a prerequisite to modeling and computation.
We circumvent this step using the recently-developed
{\em equation-free} (EF) framework for complex, multiscale systems
modeling~\cite{pnas00,review02,smallmanifesto}.
In this framework we can perform system-level computational
tasks without explicit knowledge of the coarse-grained 
equations; these unavailable equations are {\em solved}
by designing, performing and processing the results of
short bursts of appropriately initialized detailed
(fine-scale, microscopic) simulations.

We consider a paradigmatic model of coupled oscillators,
the Kuramoto model, consisting of a population of $N$ all-to-all,
phase-coupled limit-cycle oscillators with
i.i.d. $\omega$ with distribution function $g(\omega)$.
This model has been extensively studied because of its simplicity
and certain mathematical tractability, yet it is not merely
a toy model. It appears as a normal form for general systems
of coupled oscillators (e.g. Refs.~\cite{wiesenfeld96,kiss02}).

We choose a Gaussian with standard deviation $\sigma_{\omega} = 0.1$
for $g(\omega)$; however, our approach is not
limited to this particular choice, nor to the Kuramoto model.
Due to rotational symmetry, the mean frequency
$\Omega = \sum_i \omega_i/N$ can be set to 0 without loss of
generality.
The governing equation for the phase angle of the $i$th oscillator
$\theta_i$ is
\begin{equation}
\label{coupledODE}
{d\theta_i\over dt} = \omega_i+{K\over N}\sum_{j = 1}^{N}\sin(\theta_j-\theta_i),~~~~~1 \leq i \leq N,
\end{equation}
where $K \geq 0$ is the coupling strength.
Spontaneous synchronization (phase-locking) occurs at
sufficiently large $K$.
As $K$ decreases across a critical value $K_{tp}$, more and
more oscillators desynchronize until they all essentially
evolve with their own frequencies below another critical
value $K_c$~\cite{kuramoto75,strogatz00,mikhailov04}.
Kuramoto~\cite{kuramoto75} introduced a complex order parameter
$re^{i\psi} = {1 \over N}\sum_{j=1}^{N}e^{i\theta_j}$
to describe the long-time states;
the effective radius $r(t)$ measures the phase coherence;
see also Ref.~\cite{daido94} for an order {\em function}.
%
The asymptotic value of $r~(t\rightarrow \infty)$ in the
continuum-limit ($N \rightarrow \infty$) exhibits a temporal
analog of phase transition at $K_c$~\cite{kuramoto75}.

\begin{figure}[t]
\begin{center}
\includegraphics[width=.65\columnwidth]{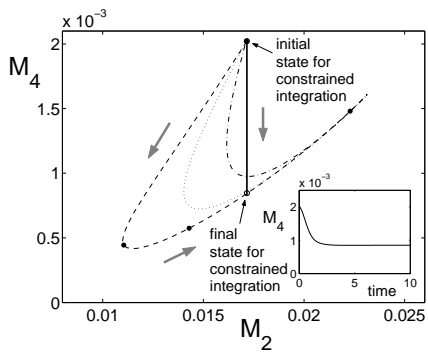}
\caption{
\label{maturation}
Three {\it coarsely identical} (same
${\cal M}_2$ and ${\cal M}_4$) but microscopically
different initializations (dashed, dotted and dot-dashed
lines; see text) evolve along {\it different trajectories},
to a slow manifold and, ultimately, the same synchronized state
($N = 300$; $K = 0.7 > K_{tp}$).
Constraining the evolution to ${\cal M}_2 = 0.017$
(solid line) guides the trajectory directly to
this slow manifold; the inset shows ${\cal M}_4$ becoming 
slaved to ${\cal M}_2$ during this constrained evolution
by $t \approx 2.0$.
}
\end{center}
\end{figure}

The order parameter $r$ conveniently represents statistical
behaviors around the critical point $K = K_c$;
however, $r$ does not uniquely specify the microscopic state,
and it may not adequately describe transient dynamics.
%
The statistical moments of the phase angle distribution function 
$
{\cal M}_n \equiv
{1\over N}\sum_{j=1}^{N}\left(\theta_j-{1\over N}\sum_{i=1}^{N}\theta_i\right)^n,
$
where $n$ is a positive integer, are a ``natural'' 
first choice of coarse-grained
observables (in a kinetic theory-like description).
Due to the symmetry, we consider only even-order moments,
and test whether a closure in terms of ${\cal M}_2$ and
${\cal M}_4$ is likely for $K \geq K_{tp}$.
We prepare several distinct initial phase angle distributions
with identical coarse-grained values (${\cal M}_2 = 0.017$
and ${\cal M}_4 = 0.0020$); these phase angles are randomly
assigned to oscillators.
%
The phase portrait in Fig.~\ref{maturation} shows direct
simulation [using Eq.~(\ref{coupledODE})] for three of these
initial assemblies; the trajectories are clearly distinct,
suggesting that the dynamics cannot close simply on these
two observables.
Including higher order moments (such as ${\cal M}_6$) 
as observables does not remedy the situation.
It is also clear, however, that the {\it long-term dynamics}
lie on a low-dimensional manifold (ultimately a one-dimensional
one) towards which all trajectories are eventually attracted.

\begin{figure}[t]
\begin{center}
\includegraphics[width=.72\columnwidth]{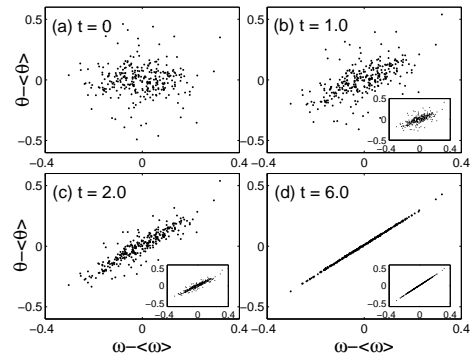}
\caption{
\label{alignment}
Time snapshots in the $\omega-\theta$ plane for free
evolution (main panels; dashed line in Fig.~\ref{maturation}) 
and for constrained evolution (insets; solid line
in Fig.~\ref{maturation}) respectively.
Each dot represents an oscillator, and
(a) to (d) are snapshots at t = 0, 1.0, 2.0, and 6.0,
respectively, marked by filled circles in Fig.~\ref{maturation}.
Strong correlations develop during the initial transient
stages (``oscillator sorting'').
}
\end{center}
\end{figure}

The dynamical differences among the three cases arise from
the microscopic differences of the (macroscopically identical)
initial conditions; this is best seen in the $\omega$-$\theta$
plane (Fig.~\ref{alignment}).
{\it Correlations} between $\theta$ and $\omega$ develop
(the initial ``cloud'' in the $\omega$-$\theta$
plane quickly evolves to a ``curve''), as all transients
initially approach the slow manifold: The oscillators
``sort themselves out''.
These correlations were not accounted for when we assigned
angles randomly to oscillators in the assembly.

We now include a ``remedial initialization'' step,
evolving the dynamics by {\it constraining them} on prescribed
values of the moments, as a system of differential
algebraic equations (DAEs) using Lagrange multipliers.
The solid line in Fig.~\ref{maturation} shows this preparatory
step with a constraint on ${\cal M}_2$ only;  constrained
evolution brings the assembly down to the right point on
the slow manifold, and the same ``sorting'' develops as in
the aforementioned freely-evolving cases.
Phase angle statistics {\em alone} do not, therefore, constitute
good observables~\cite{moon_part1};
$\omega$-$\theta$ correlations {\em should} be accounted for
in the coarse description.

Motivated by this observation, we explore the long-term
dynamics with a {\it different} set of observables,
treating both $\theta$ and $\omega$ as random variables.
The former is expanded in Hermite polynomials of the latter,
Gaussian random variable; Wiener polynomial chaos is the
appropriate choice for Gaussian distributions. 
Generalized
polynomial chaos (gPC)~\cite{xiu02} would be invoked for
different frequency distributions (e.g. we also successfully
used Legendre polynomial expansions for uniform $g(\omega)$).
For convenience, we introduce the normalized random variable
$\xi \equiv \omega/\sigma_{\omega}$:
\begin{equation}
\label{ortho}
\theta(\omega,t) =
\sum_{i=0}^p \alpha_i(t) H_i\left(\xi\right)
= \sum_{i=0}^p {\left< \theta, H_i\right> \over
\left< H_i, H_i\right>} H_i\left(\xi\right),
\end{equation}
where $p$ is the highest order retained in the truncated series,
$\left<\cdot~,\cdot\right>$ denotes the inner product with respect
to the Gaussian measure, and $H_i$ is the $i$th Hermite polynomial
[$H_0(x) = 1, H_1(x) = x, H_2(x) = x^2-1,
H_3(x) = x^3-3x, \cdots$].
Only odd-order $\alpha_i$'s are considered, due to symmetry.
We will see that here the first two nonvanishing coefficients
$\alpha_1$ and $\alpha_3$ provide an adequate representation.
Given a particular detailed realization of the oscillator state,
its PC coefficients $\alpha_i$'s are estimated
through a least squares fitting algorithm, interpreting
$\theta$ as an empirical function
$f(\xi) \equiv \alpha_1 \xi + \alpha_3 \left(\xi^3-3\xi\right)$
and minimizing the residual
$R^2 \equiv \sum_j \left[\theta_j - f(\xi_j;\alpha_1,\alpha_3) \right]^2$.
This procedure corresponds to the {\it restriction} (fine to coarse)
step in the EF framework, described below.
 
\begin{figure*}[ht]
\begin{center}
\includegraphics[width=.57\columnwidth]{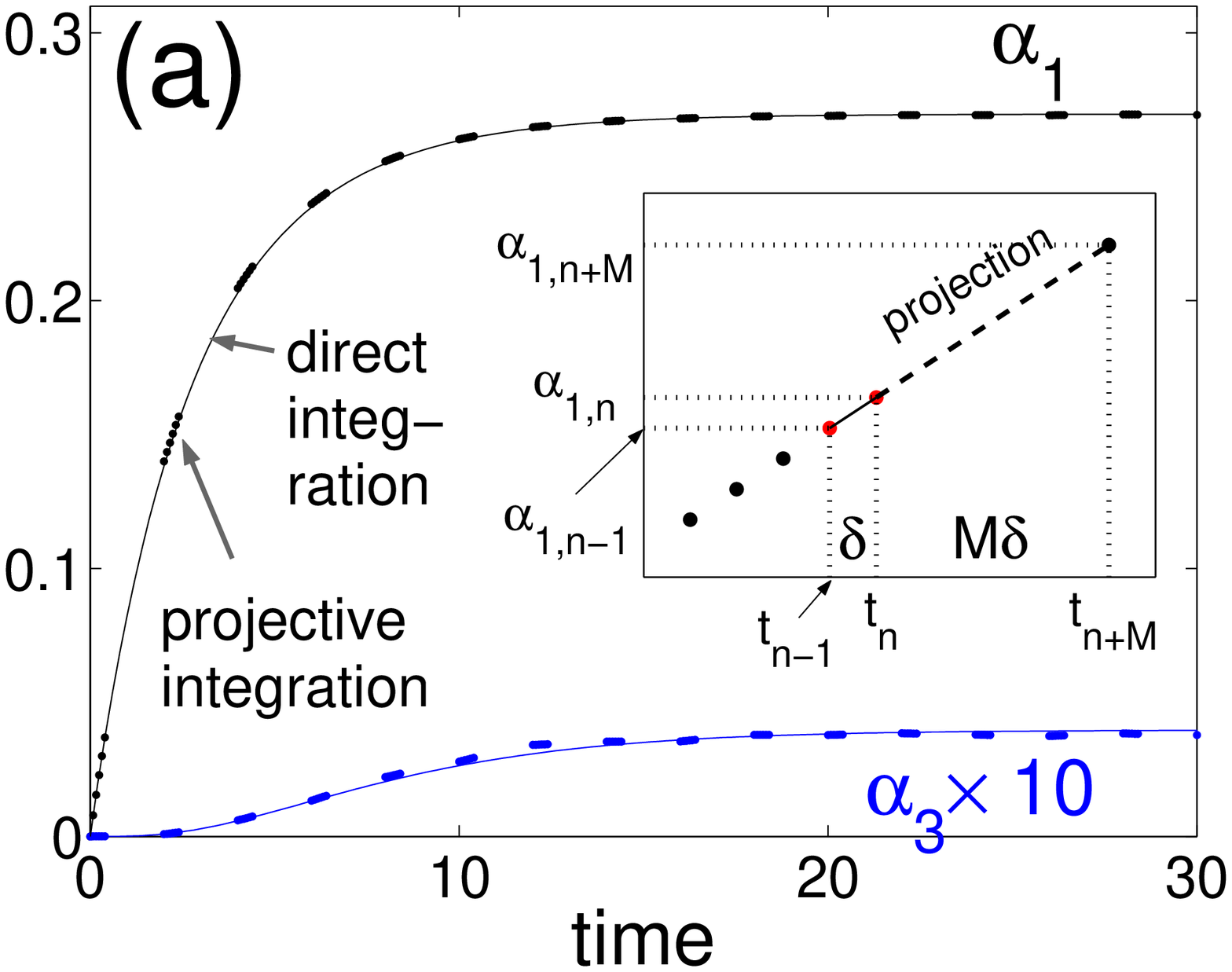}
\includegraphics[width=.58\columnwidth]{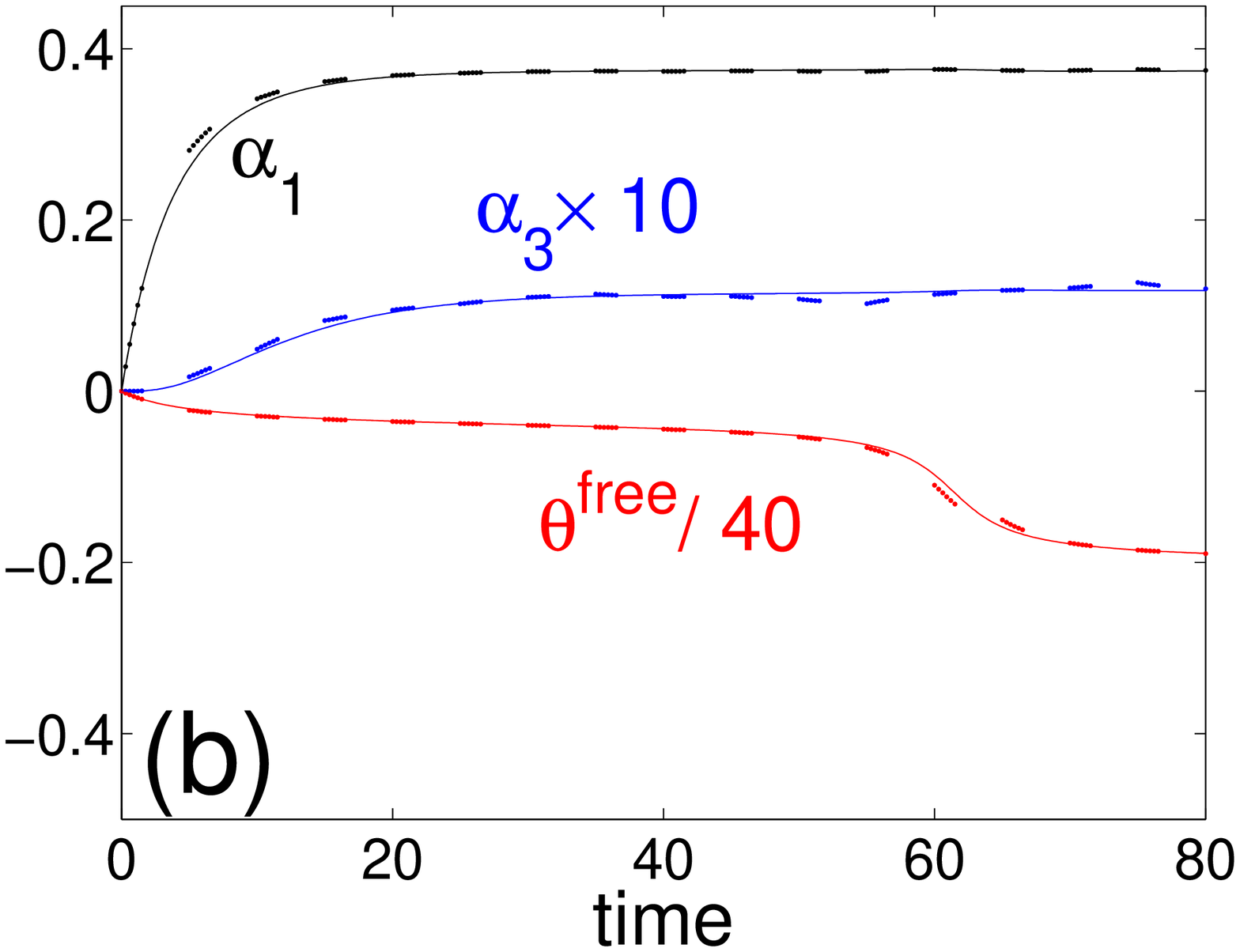}
\includegraphics[width=.58\columnwidth]{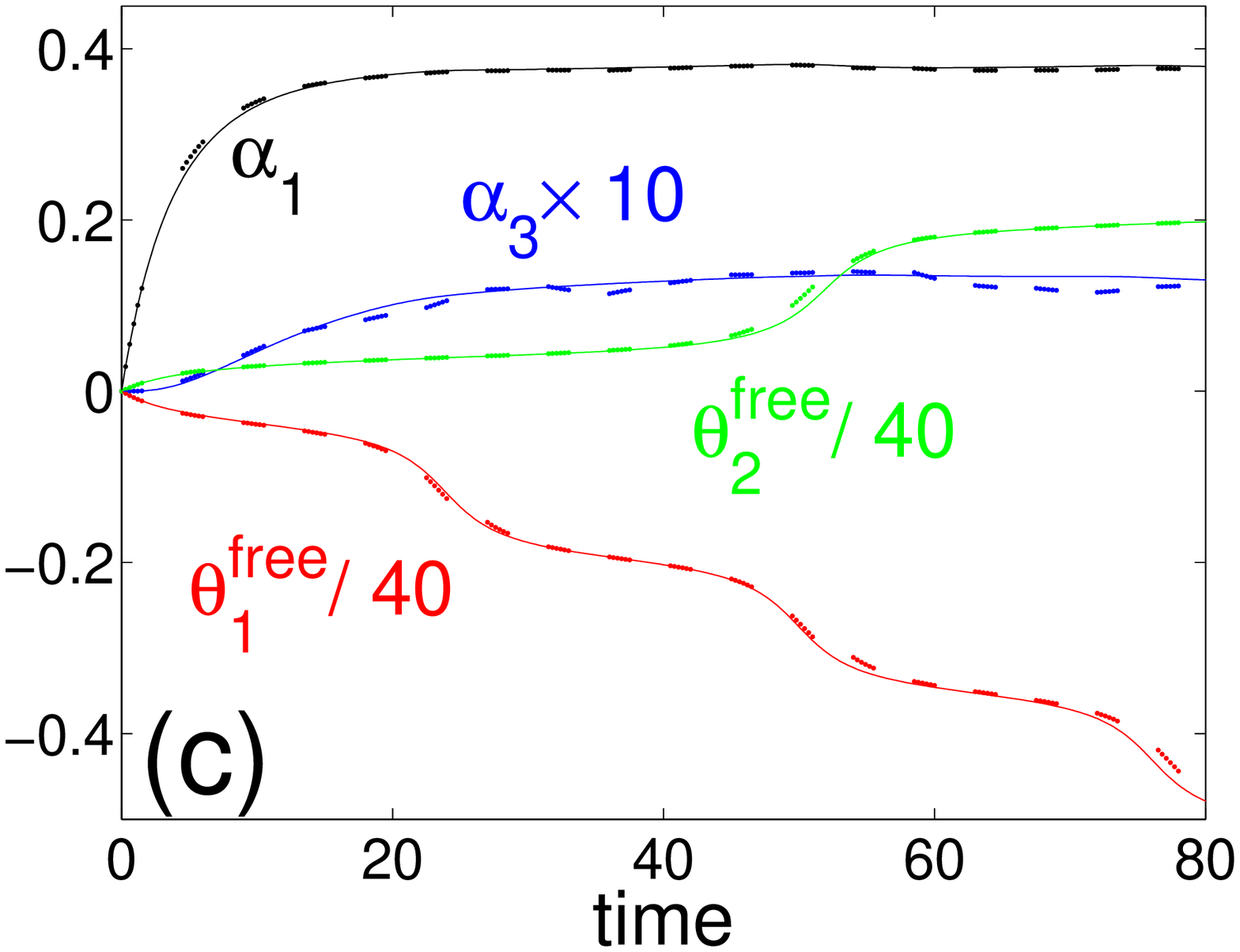}
\caption{
\label{proj_integ}
(color online)
Coarse projective integration (dots) and detailed coupled
oscillator dynamics (lines); $N = 300$.
(a) Two PC coefficients ($K = 0.4$; full synchronization).
(b) Two PC coefficients and a single ``free'' oscillator
($K = 0.31$).
(c) Two PC coefficients and two ``free" oscillators
($K = 0.31$).
Natural frequencies are newly drawn from $g(\omega)$ at each
lifting step (see text).
Inset in (a): Schematic of a projective integration step:
The last part (last two dots, at $t_{n-1}$ and $t_n$) of a short burst of 
direct integration (five dots)
is used to estimate the local time derivative (solid line).
PC values at a {\it future} time $t = t_{n+M}$
are ``projected" through forward Euler,
i.e., $\alpha_1|_{t=t_{n+M}} = \alpha_1|_{t=t_n} +
{\alpha_{1,n}-\alpha_{1,n-1} \over t_n-t_{n-1}}(t_{n+M} - t_n)$.
}
\end{center}
\end{figure*}

In the EF approach, appropriately initialized short
bursts of detailed, fine-scale simulation are used to estimate
quantities pertaining to the evolution of the coarse-grained
variables (observables).
Lacking an explicit coarse-grained model in terms of the
first few PC coefficients, we {\it estimate}
the quantities necessary for scientific computation with
it (time derivatives, action of Jacobians, residuals) 
through {\it on demand} numerical experimentation with 
the detailed, fine-scale model [Eq.~(\ref{coupledODE})].

The general procedure consists of
({\it i}) identifying good observables that sufficiently
describe the coarse-grained dynamical state (here, a few $\alpha_i$'s),
({\it ii}) constructing a {\it lifting} operator, mapping the
coarse description to one (or more, for variance reduction
purposes) consistent fine-scale realization(s) 
[randomly drawing $\omega$ from $g(\omega)$ and assigning
$\theta$, using Eq.~(\ref{ortho}) and given $\alpha_i$ values],
({\it iii}) evolving the lifted, fine-scale initial conditions
for certain time horizon,
({\it iv}) {\it restricting} the resulting fine-scale description
to the coarse observables [finding the PC coefficients of
the final state], and
({\it v}) repeating the procedure as necessary to  perform specific
scientific computation steps.
This is a general approach that has been combined with various
fine-scale models~\cite{pnas00,examples};
see Refs.~\cite{review02,smallmanifesto}.

We first demonstrate {\em coarse} projective integration~\cite{gear03}.
Each group of five dots in Fig.~\ref{proj_integ} represents
the time horizon during which the detailed equations are 
integrated to enable the projective step; 
the local time derivatives of the observables are estimated
here simply from the last two dots in each group.
Coarse variable values at a projected, future time are
estimated using these derivatives and (for projective
forward Euler) {\it linear} extrapolation in time
[see the inset in Fig.~\ref{proj_integ} (a)].
After the projection step we {\it lift} the coarse variables
to consistent fine-scale realizations, and use these
as the initial condition for another short burst of direct
detailed integration [steps ({\it ii}) and ({\it iii}) above].
Depending on the relative lengths of the projection step
($M\delta$) and the short run required to estimate the coarse time
derivatives ($n\delta$), this procedure may significantly accelerate
the computational evolution of the oscillator {\it statistics};
the cost of the lifting step (here negligible) must also be considered.
At each lifting step, $\omega$ was {\em newly} drawn from $g(\omega)$,
and the full integration
(lines) and projective integration (dots) agree on the level of
fluctuation among realizations.
The PC coefficients display smooth
behavior, nearly independently of particular random draws;
for {\it the same} random draw at every step, results would
be even better.
Projective integration in Fig.~\ref{proj_integ} (a)
reduced the computational effort in our illustrative
direct integration by a factor of four.
The numerical analysis of projective integration (stability,
accuracy, stepsize selection and estimation issues) is a topic
of current research (see e.g., Refs.~\cite{review02,gear03,ramiro04});
here we simply demonstrated the procedure and its potential.

Slightly below the transition value $K_{tp}$, where only few
oscillators become desynchronized, we consider the system as
a combination of synchronized ``bulk'' and a few ``free''
oscillators.
Good coarse-grained observables then are a few PC
coefficients for the ``bulk'' synchronized oscillators
and the phase angle(s) of the (few) desynchronized one(s).
The EF approach can be directly ``wrapped around''
this alternative representation.
Both for the one free and two free oscillator cases, projective
integrations on the new observables successfully track (and accelerate)
direct detailed simulations [Figs.~\ref{proj_integ} (b) and (c)].
These ``good observables'' are suggested by direct inspection 
and common sense; for more complicated, high-dimensional problems,
good state parameterizations require modern data mining algorithms.
Diffusion maps on graphs constructed by the data~\cite{nadler05} 
are a promising tool for detecting good ``reduction coordinates''
(observables) on which to base EF computations.

\begin{figure}[t]
\begin{center}
\includegraphics[width=.62\columnwidth]{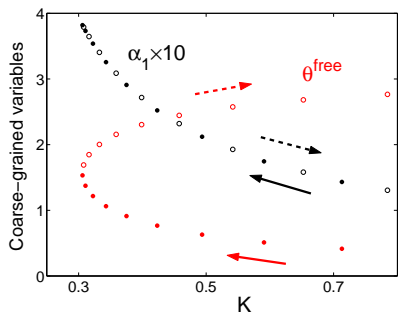}
\caption{
\label{diagram}
(color online)
Coarse bifurcation diagram for the full synchronization
regime ($K \geq K_{tp}$), obtained through {\em coarse}
Newton-GMRES method and pseudo arc-length continuation ($N = 300$).
The same variables as in Fig.~\ref{proj_integ} (b) are used;
the phase angle of the single ``free'' oscillator ($\theta^{free}$)
is an extra observable (its natural frequency
is positive in this case). 
The PC coefficients,
obtained by discounting the ``free'' oscillator, exhibit nearly the
same values both for the stable (filled circles) and the unstable (open
circles) branch (only $\alpha_1$ is shown here).
Only $\theta^{free}$ shows significant variation along the two
branches. Arrows are included to guide the eye.
}
\end{center}
\end{figure}

Direct, long-time simulation is often inefficient in computing
long-time (stationary) states.
Numerical bifurcation algorithms, more appropriate for 
stability and parametric analysis, can be implemented
in an equation-free framework: The residual
and the action of the unavailable Jacobian are numerically
estimated through short bursts of appropriately initialized
detailed simulations.
Starting from a coarse-grained initial condition, we lift,
and integrate the full model for time ${\Delta T}$.
We then restrict to the observables of the final state
$\Phi_{\Delta T}$; this is the {\it coarse time-stepper}. 
We now {\em solve} for the fixed point satisfying
$
\Phi_{\Delta T}
\left( \begin{array}{c}
\mbox{\boldmath $\alpha$}
\\
\mbox{\boldmath $\theta$}^{free}
\end{array} \right)
-
\left( \begin{array}{c}
\mbox{\boldmath $\alpha$}
\\
\mbox{\boldmath $\theta$}^{free}
\end{array} \right)
=0,
$
using the {\em coarse} Newton-GMRES~\cite{gmres}, 
a matrix-free iterative method (together with the pseudo
arc-length continuation); additional coarse
observables $\mbox{\boldmath $\theta$}^{free}$
are appended when necessary.
We construct bifurcation diagrams like the one in 
(Fig.~\ref{diagram}) with respect to the parameter $K$, showing
a turning point (actually, a ``sniper") bifurcation at $K = K_{tp}$.
A single oscillator (whose phase angle $\theta^{free}$ is
treated as a separate coarse observable) becoming ``free" from
the synchronized ``bulk" at that point.
For sufficiently large $K$ values (when $r \approx 1$) analytical
estimates of certain elements of the shape of the $\omega$-$\theta$
correlation become  possible (e.g., from Eq.~(\ref{coupledODE})
one can obtain, at steady state,
$\alpha_1 \approx  \sigma_{\omega}/K$, in reasonable agreement
with our steady state computations at $K >\sim 0.5$).

In summary, the EF multiscale approach was successfully
used for coarse-grained dynamic computations of finite
assemblies of non-identical coupled oscillators; the derivation
of explicit closures at- and close to the synchronization regime
was circumvented.
Initial transient ``sorting'' of the oscillators,
establishing correlations between natural frequencies and phase
angles, suggested Wiener PC coefficients as the
appropriate coarse observables.
If the problem dynamics can be coarse-grained,
traditional numerical analysis algorithms
can be used as protocols for the ``intelligent'' design of short bursts of
computational experiments with the detailed, fine-scale model.
The approach can be directly generalized to analyze the
simulation and modeling of more complicated oscillator dynamics.

This work was supported by DARPA and by the National Science Foundation.

\end{document}